\documentstyle[sprocl,cite]{article}

\input{psfig}
\def\be{\begin{equation}}
\def\ee{\end{equation}}
\def\bea{\begin{eqnarray}}
\def\eea{\end{eqnarray}}

\begin{document}

\title{ASTROPHYSICAL SOURCES OF HIGH ENERGY NEUTRINOS}

\author{ R.J. PROTHEROE }

\address{Department of Physics and Mathematical Physics\\
The University of Adelaide, Adelaide, Australia 5005}

\maketitle\abstracts{
\begin{center}
ABSTRACT\\
\end{center}
I give a brief critical review of the predicted intensity of
diffuse high energy neutrinos of astrophysical origin over the
energy range from $\sim 10^{12}$ eV to $\sim 10^{24}$ eV.
Neutrinos from interactions of galactic cosmic rays with
interstellar matter are guaranteed, and the intensity can be
reliably predicted to within a factor of 2 up to $10^{17}$ eV.
Somewhat less certain are intensities in the same energy range
from cosmic rays escaping from normal galaxies or active galactic
nuclei (AGN) and interacting with intracluster gas.  At higher
energies, neutrinos will be produced by interactions of
extragalactic cosmic rays with the microwave background.  Other
sources, such as AGN, in particular blazars, and topological
defects, are more speculative.  However, searches for neutrinos
from all of these potential sources should be made because their
observation would have important implications for high energy
astrophysics and cosmology.}

\section{Introduction}

The technique for constructing a large area (in excess of $10^4$
m$^2$) neutrino telescope has been known for more than a decade
\cite{BerZat77}.  The pioneering work of the DUMAND Collaboration
developed techniques to instrument a large volume of water in a
deep ocean trench with strings of photomultipliers to detect
Cherenkov light from neutrino-induced muons.  Locations deep in
the ocean shield the detectors from cosmic ray muons.  The DUMAND
detector \cite{Learned91} was designed to be most sensitive to
neutrinos above about 1 TeV, and prototypes of other similar
experiments are already in operation such as that in Lake Baikal,
Siberia \cite{Baikal}, and NESTOR off the coast of Greece
\cite{NESTOR}.  An exciting recent development has been the
construction of a DUMAND type detector deep in the polar ice cap
at the South Pole.  This experiment called AMANDA uses the same
principle as DUMAND but takes advantage of excellent transparency
of the polar ice under extreme pressures and a stable environment
in which to embed the detectors \cite{AMANDA}.  These experiments
have the potential to be expanded in the future to a detector on
the 1 km$^3$ scale, and a workshop was held in Aspen in 1996 to
consider this possibility, and decide on the most suitable energy
ranges for observation.  It is therefore appropriate and timely
to review the predicted intensity of astrophysical sources of
neutrinos over a the entire energy range from $\sim 10^{12}$ eV
to $\sim 10^{24}$ eV.

\section{Neutrinos from Cosmic Ray Interactions}

\begin{figure}[htb]
\psfig{figure=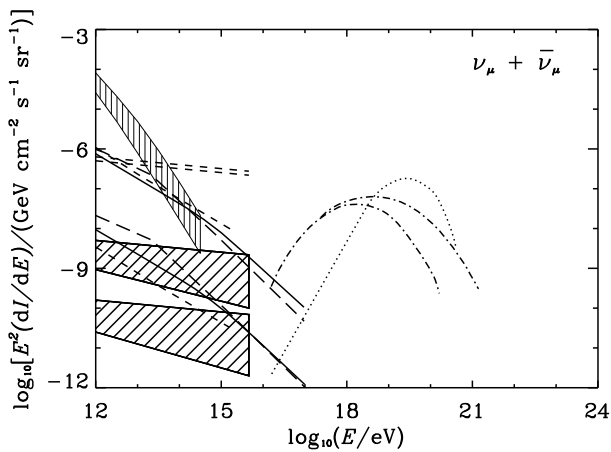,height=2.5in}
\caption{Neutrinos from cosmic ray interactions with the interstellar medium
(upper curves for $\ell=0^\circ, \; b=0^\circ$, lower curves for $b=90^\circ$):
-- -- -- Domokos et al. \protect\cite{Domokos};
- - - - Berezinsky et al. \protect\cite{Berezinsky93};
---------- Ingelman and Thunman \protect\cite{IngelmanThunman96}.
The band with vertical hatching shows the range of atmosheric neutrino
background as the zenith angle changes \protect\cite{Lipari}. 
Neutrinos from cosmic ray interactions with the microwave background:
$- \cdot - \cdot - \cdot -$ Protheroe and Johnson \protect\cite{ProtheroeJohnson95};
$\cdots \cdots$ Hill and Schramm \protect\cite{Hil85}.
Neutrinos from cosmic ray interactions in clusters of galaxies
(Berezinsky et al. \protect\cite{BerezinskyBlasiPtuskin96}):
lower hatched area -- normal galaxies; upper hatched area -- AGN; 
= = = = = upper limit if diffuse $\gamma$-ray flux originates in clusters.
\label{fig:cr}}
\end{figure}

There will definitely exist an important galactic diffuse
neutrino background due to interactions of the galactic cosmic
rays with interstellar matter.  The spectrum of galactic cosmic
rays is reasonably well known, as is the matter distribution in
our galaxy.  Estimates of the intensity have been made by Domokos
et al\cite{Domokos}, Berezinsky et al. \cite{Berezinsky93}, and
Ingelman and Thunman \cite{IngelmanThunman96} and their
predictions are shown in Fig.~\ref{fig:cr}.  The differences of
about a factor of 2 between the predictions are accountable in
terms of the slightly different models of the interstellar matter
density, and cosmic ray spectrum and composition used.  Also
shown is the atmospheric neutrino background as estimated by
Lipari \cite{Lipari}.  In addition, there will be a very
uncertain background (not plotted) due to charm production (see
Gaisser et al. \cite{GaisserHalzenStanev95} for a survey of
predictions).

Moving to higher energies, cosmic rays above $\sim 10^{20}$ eV
will interact with photons of the cosmic microwave background
radiation \cite{Gre66,Zat66}.  Again, we know that both
ingredients exist (the two highest energy cosmic rays detected
have energies of $3 \times 10^{20}$ eV \cite{Bir95} and $2 \times
10^{20}$ eV \cite{Hay94}), and so the pion photoproduction at
these energies will occur resulting in a diffuse neutrino
background.  However, the intensity in this case is
model-dependent because it is not certain precisely what is
origin of the highest energy cosmic rays, and indeed if they are
extragalactic, although this seems very probable (see my first
lecture \cite{ProtheroeERICE96CR} for a discussion of the highest
energy cosmic rays).  One of the most likely explanations of the
highest energy cosmic rays is acceleration in Fanaroff-Riley
Class II radio galaxies as suggested by Rachen and Biermann
\cite{RachenBiermann93}.  Protheroe and
Johnson\cite{ProtheroeJohnson95} have repeated Rachen and
Biermann's calculation in order to calculate the flux of diffuse
neutrinos and $\gamma$-rays which would accompany the UHE cosmic
rays, and their result has been added to Fig.~\ref{fig:cr}.  Any
model in which the cosmic rays above $10^{20}$ eV are of
extragalactic origin will predict a high energy diffuse neutrino
intensity probably within an order of magnitude of this at
$10^{18}$ eV.  For example, I also show an earlier estimate by
Hill and Schramm \cite{Hil85}.

Somewhat less certain is the flux of neutrinos from clusters of
galaxies.  This is produced by $pp$ interactions of high energy
cosmic rays with intracluster gas.  Berezinsky et
al. \cite{BerezinskyBlasiPtuskin96} have recently made
predictions of this, and I have added to Fig.~\ref{fig:cr} their
estimates of the diffuse neutrino intensity due to interactions
of cosmic rays produced by normal galaxies and AGN.  Also shown
is their upper limit estimated assuming that all the diffuse
$\gamma$-ray background is due to such interactions in clusters.
However, this is unlikely to be the case (see next section).

\section{Neutrinos from Active Galactic Nuclei}

The second EGRET catalog of high-energy $\gamma$-ray sources
\cite{Thompson95} contains 40 high confidence identifications of
AGN, and all appear to be blazars.  Clearly, the $\gamma$-ray
emission is associated with AGN jets.  Blazars appear also to be
able to explain the bulk of the diffuse $\gamma$ ray emission
\cite{Chiang95}, and models where $\gamma$-ray emission does not
originate in the jet are unlikely to contribute significantly to
the diffuse $\gamma$-ray (and neutrino) intensity (see Protheroe
and Szabo \cite{ProtheroeSzabo94} and references therein for
predictions for non-blazar AGN).  Several of the EGRET AGN show
$\gamma$-ray variability with time scales of $\sim 1$ day
\cite{Kniffen93} at GeV energies, and variability on time scales
of $\sim 1$ hour or less \cite{Gaidos96} has been observed at
TeV.  This variability places an important constraint on the
models, and not all models developed so far are consistent with
this.  I shall survey the neutrino emission predicted in blazar
models irrespective of this , optimistically assuming these
models may be made to accommodate the latest variability
measurements.

Most theoretical work on $\gamma$-ray emission in AGN jets
involved electron acceleration and inverse Compton scattering,
and these models will predict no neutrinos.  The alternative
approach is to accelerate protons instead of, or as well as,
electrons.  In this case interactions of protons with matter or
radiation would lead to neutrino production.  In some of the
proton models energetic protons interact with radiation via pion
photoproduction (see my first lecture \cite{ProtheroeERICE96CR}
for a discussion of $p \gamma$ interactions).  This radiation may
be reprocessed or direct accretion disk radiation
\cite{Protheroe96}, or may be produced locally, for example, by
synchrotron radiation by electrons accelerated along with the
protons \cite{MannheimBiermann92,Mannheim95}.  Pair synchrotron
cascades initiated by photons and electrons resulting from pion
decay give rise to the emerging spectra, and this also leads to
quite acceptable fits to the observed spectra.  This approach has
the obvious advantage of leading to potentially much higher
photon energies, because protons have a much lower synchrotron
energy loss rate than electrons for a given magnetic environment.
In both classes of model, shock acceleration has been suggested
as the likely acceleration mechanism (see my first lecture
\cite{ProtheroeERICE96CR} for a discussion of and references to
shock acceleration).

\begin{figure}[htb]
\psfig{figure=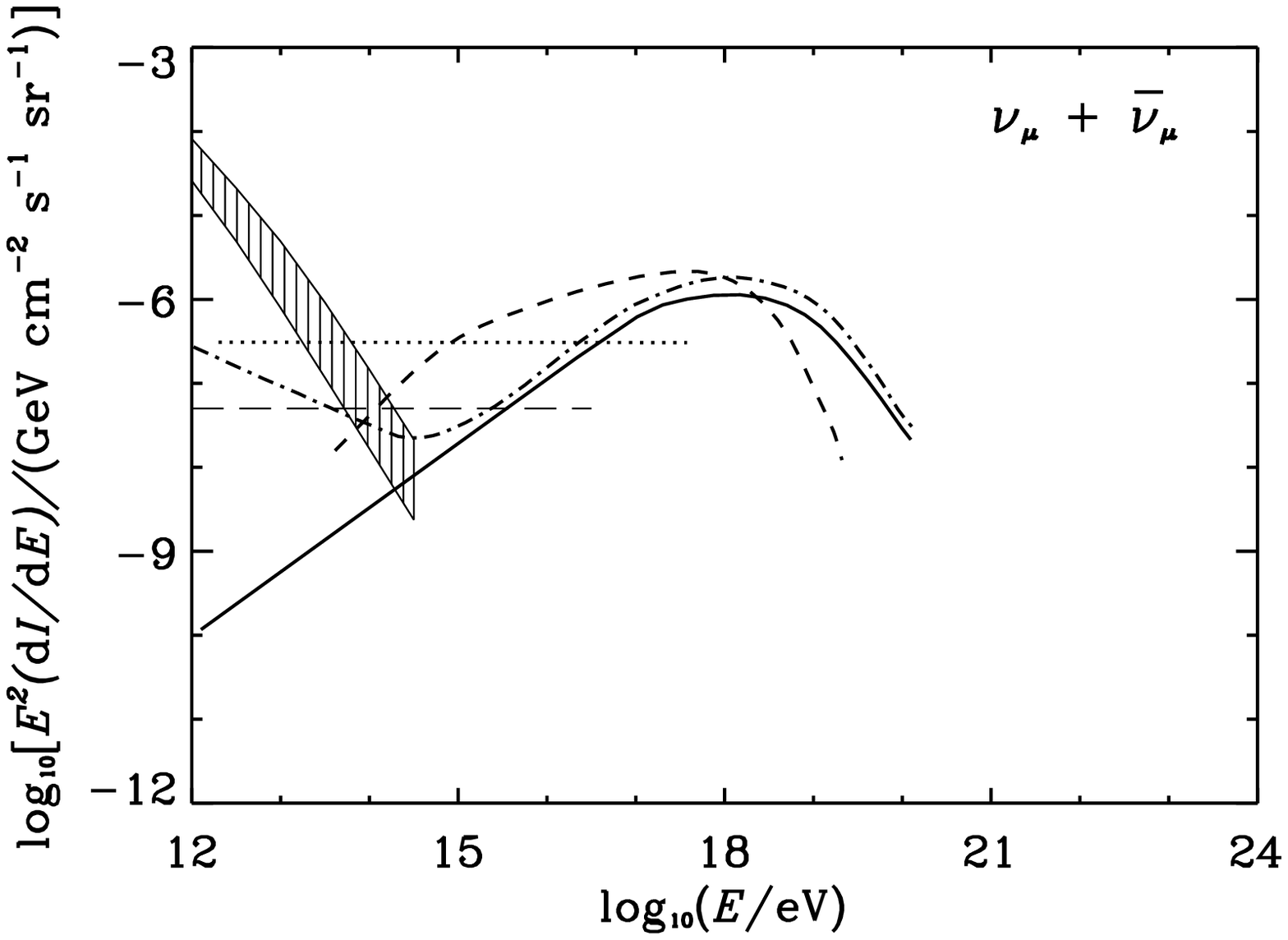,height=2.5in}
\caption{ Diffuse neutrinos from blazars:
---------- $p \gamma$ Mannheim \protect\cite{Mannheim95}; 
$- \cdot - \cdot - \cdot -$ $pp$ and $p \gamma$ 
Mannheim \protect\cite{Mannheim95}; 
- - - - - $p \gamma$ Protheroe \protect\cite{Protheroe96}; 
--- --- --- --- Stecker \protect\cite{SDSS93}; 
$\cdots \cdots$ Nellen \protect\cite{Nellen}.
\label{fig:bl}}
\end{figure}

By appropriately integrating over redshift and luminosity in an
expanding universe, using a luminosity function (number density
of objects per unit of luminosity) appropriate to blazars, and
using the proton blazar models to model the $\gamma$ ray and
neutrino spectra one can estimate the diffuse neutrino background
expected from blazars.  Fig.~\ref{fig:bl} shows intensities of
$(\nu_\mu + \bar{\nu}_\mu)$ in proton blazar models estimated by
Mannheim \cite{Mannheim95} and Protheroe\cite{Protheroe96}
together with estimates for blazars obtained under very different
assumptions by Stecker \cite{SDSS93} and Nellen {\it et al.}
\cite{Nellen}.  For some of these models Hill \cite{GaryHill96}
has calculated expected muon rates.

\section{Neutrinos from Exotic Sources}

\begin{figure}[htb]
\psfig{figure=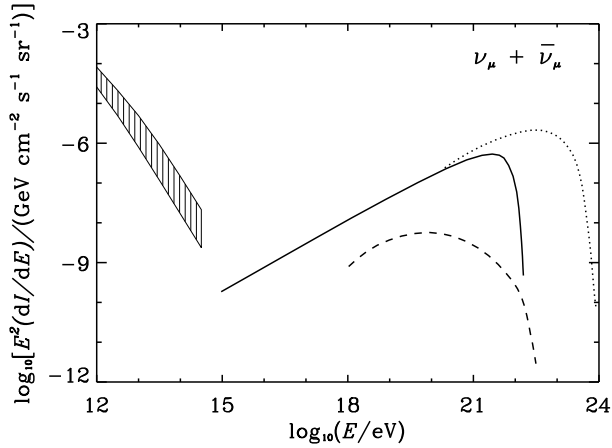,height=2.5in}
\caption{Neutrinos from exotic sources: 
--------------- TD model just ruled out according to 
Protheroe and Stanev \protect\cite{ProtheroeStanev96};
$\cdots \cdots$ TD model just allowed according to 
Sigl et al. \protect\cite{SiglLeeCoppi96};
-~-~-~-~-~-~ GRB model according to Lee \protect\cite{Lee}.
\label{fig:ex}}
\end{figure}

Finally, I discuss perhaps the most uncertain of the components
of the diffuse high energy neutrino background, that due to
exotic sources such as topological defects.  The possibility that
topological defects could be responsible for the highest energy
cosmic rays was discussed briefly in my first lecture
\cite{ProtheroeERICE96CR}.  Although these models appear to be
ruled out by the high GeV $\gamma$-ray intensity produced in
cascades initiated by X-particle decay
\cite{ProtheroeJohnsonTaup95,ProtheroeStanev96}, I plot in
Fig.~\ref{fig:ex} the neutrino emission for a set of TD model
parameters just ruled out according to Protheroe and Stanev
\cite{ProtheroeStanev96} and just allowed according to Sigl et
al. \cite{SiglLeeCoppi96}.  I emphasize that these predictions
are {\em not} absolute predictions but the intensity of
$\gamma$-rays and nucleons in the resulting cascade is normalized
in some way to the highest energy cosmic ray data.  It is my
opinion
\cite{ProtheroeJohnsonTaup95,ProtheroeStanev96,ProtheroeERICE96CR}
that these models are neither necessary nor able to explain the
highest energy cosmic rays without violating the GeV $\gamma$-ray
flux observed.  Also shown is an estimate of the diffuse neutrino
intensity estimated in a model in which the highest energy cosmic
rays have their origin in sources of gamma ray bursts \cite{Lee}

The intensities shown in Fig.~\ref{fig:ex} are {\em extremely}
uncertain.
Nevertheless, it is important to search for such emission because,
if it is found, it would overturn our current thinking on the origin 
of the highest energy cosmic rays and, perhaps more importantly,
our understanding of the universe itself.

\section{Conclusion}

It is beyond doubt that astrophysical sources of high energy
neutrinos exist.  Given that we know cosmic rays exist in our
galaxy which contains interstellar matter, cosmic ray
interactions will result in the galaxy glowing in high energy
neutrinos.  The neutrino intensities are likely to be just
observable with the 1 km$^3$ neutrino detectors currently under
consideration.  The real challenge and interest is in attempting
to detect emission from other possible astrophysical sources such
as AGN, and the emission due to cascades initiated in the
microwave background by high energy cosmic rays accelerated in
radio galaxies, or by topological defects.  However, we cannot
predict everything -- unexpected discoveries have always occurred
whenever a new window on the universe is opened and it would be
surprising if this time there were no surprises!  It is therefore
very important that neutrino astronomy with 1 km$^3$ detectors
proceed rapidly.

\section*{Acknowledgments}

This work has benefited from discussions during the workshop on
high energy neutrino astrophysics held at the Aspen Center for
Physics in 1996 organized by Tom Weiler.  I thank in particular
Paolo Gondolo, Karl Mannheim, G\"{u}nter Sigl, and Enrique Zas,
for discussions at Aspen, and Wlodek Bednarek, Gary Hill and
Qinghuan Luo for reading the manuscript.  This research is
supported by a grant from the Australian Research Council.

\end{document}